\documentclass[twocolumn,floatfix,pra,aps,superscriptaddress]{revtex4}
\usepackage{epsfig,graphicx}
\usepackage{flafter}
\usepackage{amsmath}
\usepackage{color}
\usepackage{bm}
\usepackage{amssymb}
\usepackage{epstopdf}
\usepackage{amstext}
\usepackage{amsthm}
\usepackage{amsfonts}
\usepackage{latexsym}
\usepackage{multirow}
\usepackage{mathtools}
\usepackage{bm}
\usepackage{bbm}
    \usepackage{amsmath}
    \usepackage{amsthm}
    \usepackage{amsfonts}
    \usepackage{braket}
\usepackage{soul}
\usepackage{comment}

\begin{document}

\title{Quantized vortices in dipolar supersolid Bose-Einstein condensed gases} 

\author{A. Gallem\'{\i}}
\affiliation{INO-CNR BEC Center and Dipartimento di Fisica, Universit\`a degli Studi di Trento, 38123 Povo, Italy}
\affiliation{Trento  Institute  for  Fundamental  Physics  and  Applications,  INFN,  38123,  Trento,  Italy}
\author{S. M. Roccuzzo}
\affiliation{INO-CNR BEC Center and Dipartimento di Fisica, Universit\`a degli Studi di Trento, 38123 Povo, Italy}
\affiliation{Trento  Institute  for  Fundamental  Physics  and  Applications,  INFN,  38123,  Trento,  Italy}
\author{S. Stringari}
\affiliation{INO-CNR BEC Center and Dipartimento di Fisica, Universit\`a degli Studi di Trento, 38123 Povo, Italy}
\affiliation{Trento  Institute  for  Fundamental  Physics  and  Applications,  INFN,  38123,  Trento,  Italy}
\author{A. Recati\footnote{Corresponding Author: alessio.recati@ino.it}
}
\affiliation{INO-CNR BEC Center and Dipartimento di Fisica, Universit\`a degli Studi di Trento, 38123 Povo, Italy}
\affiliation{Trento  Institute  for  Fundamental  Physics  and  Applications,  INFN,  38123,  Trento,  Italy}


\begin{abstract}

We investigate the properties of quantized vortices in a dipolar Bose-Einstein 
condensed gas by means of a generalised Gross-Pitaevskii equation. The size of the 
vortex core hugely increases by increasing the weight of the dipolar interaction and
approaching the transition to the supersolid phase. The critical angular velocity 
for the existence of an energetically stable vortex decreases in the supersolid, 
due to the reduced value of the density in the interdroplet region. The angular 
momentum per particle associated with the vortex line is shown to be smaller than 
$\hbar$, reflecting the reduction of the global superfluidity. The real-time vortex 
nucleation in a rotating trap is shown to be triggered, as for a standard condensate, 
by the softening of the quadrupole mode. For large angular velocities, when the  
distance between vortices becomes comparable to the interdroplet distance, the 
vortices are arranged into a honeycomb structure, which coexists with the triangular 
geometry of the supersolid lattice and persists during the free expansion of the atomic 
cloud.

\end{abstract}

\maketitle

\section{Introduction}
\begin{figure}[t!]
\centering
\includegraphics[width=\columnwidth]{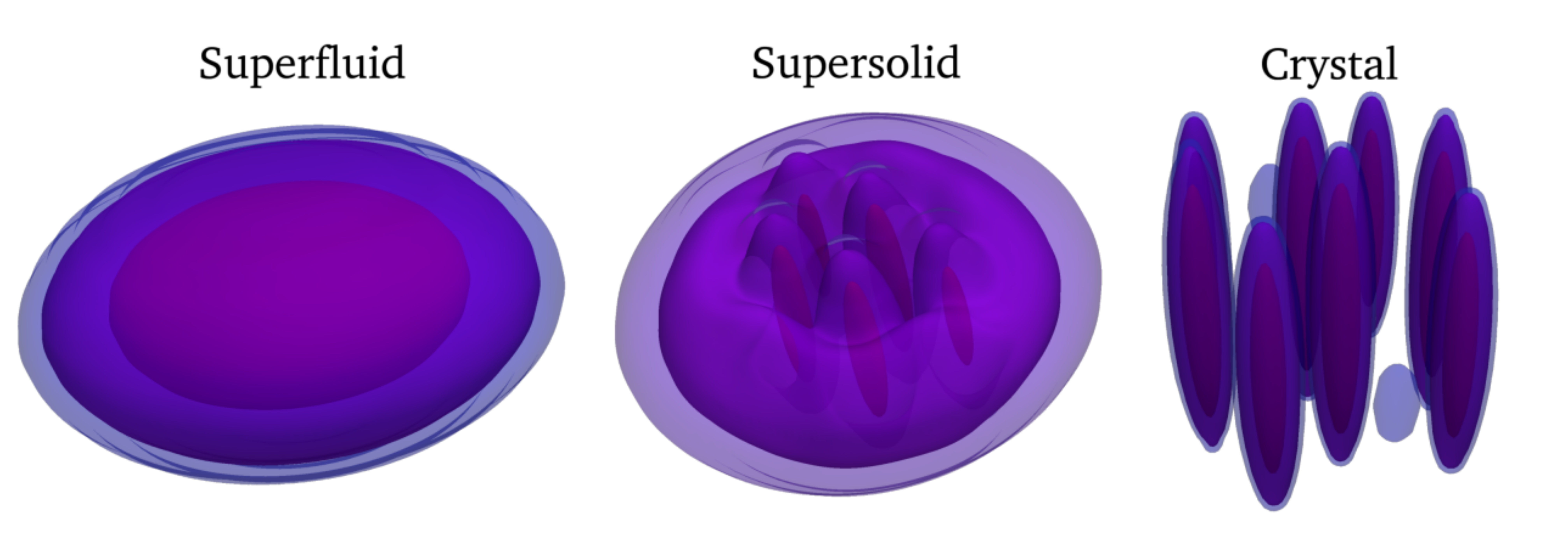}
\caption{Example of the three distinct phases of a dipolar Bose gas in a pancake geometry (from 
left to right): superfluid, supersolid and droplet crystal phase. The pictures are obtained as 
ground state solutions of the extended Gross-Pitaevskii equation (\ref{gpequation}) for $10^5$ 
atoms by decreasing the $s$-wave scattering length in order to favor the long-range anysotropic 
dipolar interaction (see text).}
\label{fig1}
\end{figure}
The recent realization of supersolidity in dipolar Bose-Einstein condensed gases~\cite{F1,S1,I1} 
is stimulating novel experimental and theoretical work aimed at studying the superfluid properties of 
these intriguing systems, which exhibit the spontaneous breaking of both gauge and translational 
symmetry yielding superfluidity and crystal periodic order,  respectively~(see, e.g., 
\cite{AndreevLifshitz,Son,Hydro-Rica2007,Roccuzzo1}). Experimental evidence of phase 
coherence among the droplets forming the crystal structure \cite{F1,I1,S1}, the occurrence 
of Goldstone modes associated with the spontaneous breaking of both symmetries \cite{F2,S2,I2} 
and the reduction of the moment of inertia with respect to the rigid value~\cite{F3} have 
provided  important signatures of the superfluid behavior of these systems. Conclusive proof of 
superfluidity is however given by the observation of quantized vortex lines, following the seminal 
papers of Refs. \cite{Cornell1999,dalibard2000,Abo-Shaeer476,Cornell2001} and \cite{Zwierlein2005} 
in Bose-Einstein condensates and strongly interacting Fermi gases, respectively. The realization of 
quantized vortices, hosted by the crystal configuration of the supersolid, then represents a 
challenging task to pursue. 
This possibility, so far not yet experimentally realized, has been the object of first 
recent theoretical investigations \cite{roccuzzo2}. Even the structure of quantized 
vortices in the fully-superfluid phase and in particular the effect of the long-range 
dipolar force on the size of the vortex core and on the value of the critical angular 
velocity needed to ensure the energetic stability of a single vortex line, represents 
an interesting topic, hopefully of near future experimental investigation. The purpose of 
this paper is to provide a first comprehensive theoretical investigation of the structure 
of quantized vortices in a Bose-Einstein condensate characterized by a long-range dipolar 
interaction with special focus on the supersolid phase. 

Our investigation is based on the use of a suitable extension of the Gross-Pitaevskii equation 
to include the beyond mean-field term (see Section~\ref{sec:GPE}) in the equation of state 
accounting for quantum fluctuations~\cite{Pelster}, which plays a crucial role in the emergence 
of supersolidity and the formation of self-bound droplets (see Fig.~\ref{fig1}). In Section 
\ref{sec:1v} we explore the properties of a single vortex line in both the superfluid and the 
supersolid case. In the superfluid phase the dipolar interaction hugely increases the value of 
the healing length when compared to Bose gases with only zero-range interaction. In the 
supersolid phase, vortices are hosted in the region separating the droplets forming the crystal 
structure and their shape is strongly deformed by the presence of the droplets. We show that the 
value of the critical angular velocity exhibits an important reduction by increasing the ratio 
between the dipolar and the zero-range strengths of the interatomic force. Furthermore we show that 
the angular momentum carried by a vortical line in an axi-symmetrically trapped supersolid is reduced 
with respect to the usual value $\hbar$ as a consequence of the reduced superfluidity of the system. 
By carrying out a time-dependent simulation we also point out that the nucleation process for the 
creation of a vortical line in the supersolid phase is favored by the softening of the quadrupole mode 
frequency. In Section~\ref{sec:lattice} we study the case of higher angular velocities and show that 
the coexistence of the density modulation and vorticity yields a honeycomb vortex lattice in place 
of the usual trangular (Abrikosov) lattice, the coexistence persisting during the expansion following 
the sudden release of the trap. We sum up our conclusions in Section~\ref{sec:conc}.


\section{Dipolar Gross-Pitaevskii equation with Lee-Huang-Yang correction}
\label{sec:GPE}

We consider a dipolar Bose gas of $^{164}$Dy atoms trapped by an in-plane isotropic 
harmonic potential $V_{\rm ho}({\bf r})=\frac{1}{2}m(\omega_x^2 x^2+\omega_x^2 y^2+\omega_z^2 z^2)$, 
with, if not differently stated, $\omega_x=\omega_y$ and $m$ the atomic mass. At zero temperature the gas 
can be characterized by a single macroscopic wave function $\Psi({\bf r},t)$, whose 
temporal evolution is described by a generalized Gross-Pitaevskii equation. The latter 
takes into account the contact, the dipole-dipole interaction and the quantum fluctuations 
\cite{Wachter2016}, and can be written as
\begin{align}
&i\hbar\frac{\partial}{\partial t}  \Psi({\bf r},t)= \mathcal{H}({\bf r})\,\Psi({\bf r},t)\,,
\label{gpequation}
\end{align}
where the hamiltonian is
\begin{align}
\mathcal{H}({\bf r})=-&\frac{\hbar^2}{2m}\nabla^2+V_{\rm ho}({\bf r})+g|\Psi({\bf r},t)|^2+\gamma(\varepsilon_{dd})|\Psi({\bf r},t)|^3\nonumber\\
+& \int d{\bf r'}V_{dd}({\bf r}-{\bf r'})|\Psi({\bf r'},t)|^2\,,
\label{gpequation2}
\end{align}
with $g=4\pi\hbar^2a/m$ the coupling constant fixed by the $s$-wave scattering length $a$ and 
$V_{dd}({\bf r}_{i}-{\bf r}_{j})=\frac{\mu_0\mu^2}{4\pi}\frac{1-3\cos^2\theta}{|{\bf r}_{i}-{\bf r}_{j}|^3}$ 
the dipole-dipole potential, being $\mu_0$ the magnetic permeability in vacuum, $\mu$ the magnetic dipole 
moment and $\theta$ the angle between the vector distance between dipoles and the polarization direction, 
which we choose as the $z$-axis. In the absence of trapping, the system can be fully characterised by the 
single parameter $\varepsilon_{dd}=\mu_0\mu^2/(3g)=a_{dd}/a$, i.e., the ratio between the strength 
of the dipolar and the contact interaction, eventually written in terms of the dipolar length $a_{dd}$ 
and the scattering length $a$. For the atom we are using, $a_{dd}=131\,a_B$, where $a_B$ is the Bohr 
radius. The third term of the Hamiltonian Eq. (\ref{gpequation2}) corresponds to the local density 
approximation of the beyond-mean-field Lee-Huang-Yang (LHY) correction 
\cite{FischerLHY,Pelster}, with 
\begin{equation}
\gamma(\varepsilon_{dd})=\frac{16}{3\sqrt{\pi}} ga^{\frac{3}{2}}\,\mbox{Re}\bigg[\!\int_0^{\pi}\!\!\!\!d\theta\sin\theta [1+\varepsilon_{dd}(3\cos^2\theta-1)]^{\frac{5}{2}}\bigg]\,.
\end{equation}
Experimental measurements and microscopic Monte Carlo calculations \cite{Saito2016} have 
confirmed that the LHY term is an accurate correction to the mean-field theory given by the 
Gross-Pitaevskii equation both in dipolar gases and quantum mixtures 
\cite{PetrovDrop,TarruellDrop,TarruellDrop2,FattoriDrop,FattoriColl}. At the mean-field level, 
increasing the role of the dipolar interaction would lead to the collapse of the cloud because 
of the attractive component of the dipolar force. The collapse is prevented by a stronger confinement 
in the polarization direction $z$, ($\omega_z>\omega_x=\omega_y$) causing the occurrence of a typical roton-like 
excitation spectrum~\cite{SantosRoton}, whose gap becomes smaller and smaller as one increases 
$\epsilon_{dd}$, and by the inclusion of the LHY term. Both effects are responsible for the emergence 
of new interesting phases. In particular a dipolar Bose gas confined in the polarization direction 
has been shown to be fully superfluid for a value of $\varepsilon_{dd}$ lower than a certain critical 
value of the order of $1.3$ (this value has a weak dependence on the trapping parameters and the 
total atom number $N$). Above this value -- which in absence 
of the LHY effect would lead to collapse due to roton softening -- 
the system presents supersolid properties characterized by density modulations and 
coherence between the density peaks. By further increasing $\varepsilon_{dd}$, the system enters the 
crystal phase, where coherence between the density peaks is destroyed and global superfluidity 
is lost.

In Ref.~\cite{roccuzzo2} we have recently shown that supersolids are able to host quantized vortices 
in the low density region between the density peaks, with a deformed vortex core. In this work we 
determine the behaviour of the relevant properties of the vortices by changing $\varepsilon_{dd}$.

In order to study vortices we add to the hamiltonian of the system the angular momentum constraint 
$-\Omega L_z$, where $L_z$ is the angular momentum operator and $\Omega$ is the angular rotation 
frequency, and solve the Gross-Pitaevskii equation either in imaginary or real time as explained in 
detail in the following. In particular, the ground state of a superfluid is insensible to rotations 
for $\Omega$ lower than a critical value $\Omega_c$, above which the presence of a vortex in the 
system becomes energetically favourable. The new stable vortical configuration carries an angular 
momentum per particle equal to $\hbar$. Supersolids, instead, react to any value of the angular 
frequency due to the existence of a nonsuperfluid component ~\cite{roccuzzo2} and for the same 
reason, as we explicitly show in the next section, the angular momentum carried by a vortex line is expected to be smaller than the usual 
value $\hbar$.

\begin{figure}[t!]
\centering
\includegraphics[width=\columnwidth]{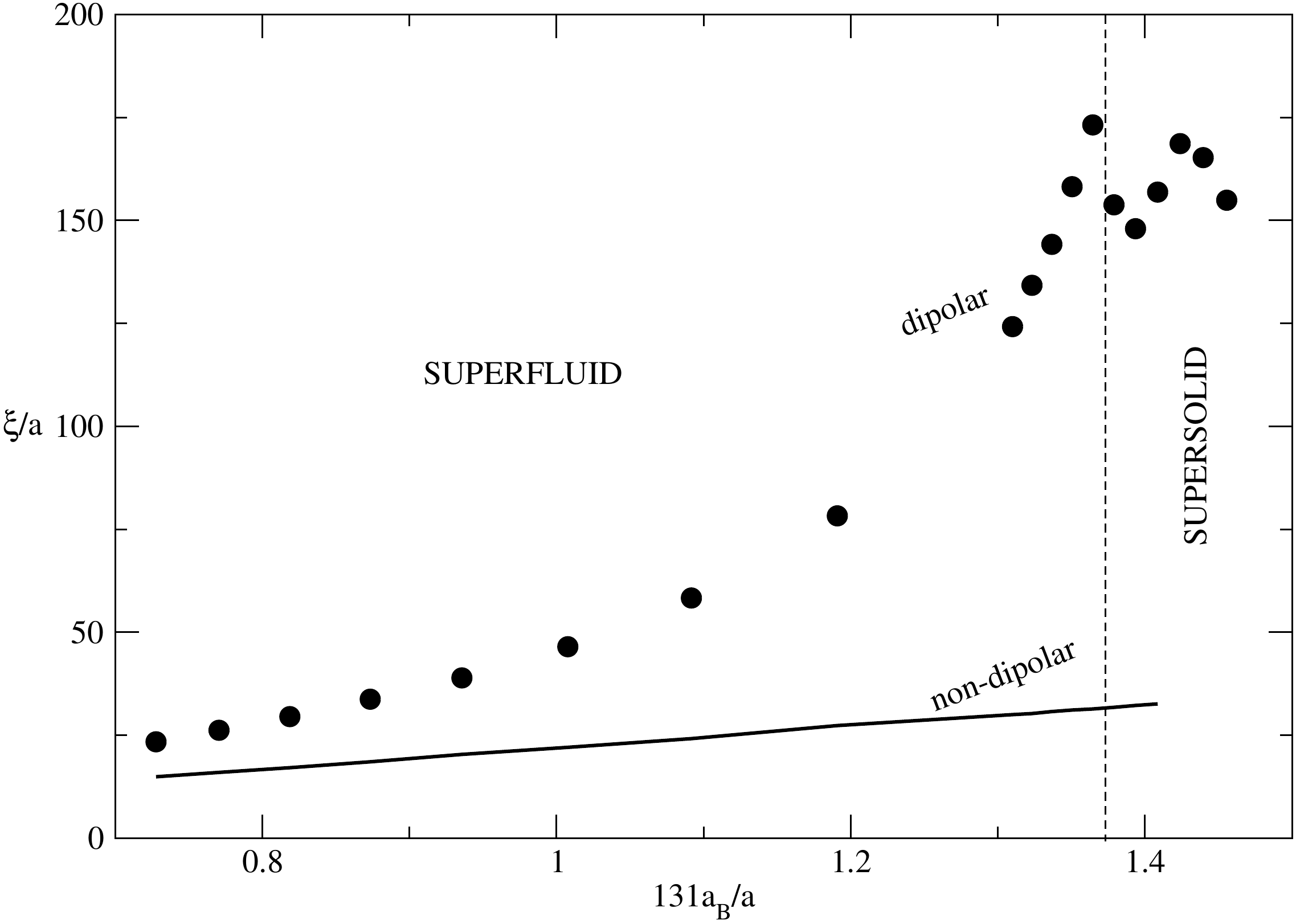}
\caption{Healing length defined as the vortex core size (see text) as a function of the scattering 
length $a$ by crossing the superfluid to supersolid transition, both in the non-dipolar (solid line) 
and dipolar (points) case. The simulation's parameters are: $N=40000$ atoms,
$\omega_{x,y,z}= 2\pi\times (60,60,120)$ Hz, and 
$\Omega=2\pi\times 12.7$ Hz.}
\label{figcore}
\end{figure}

\section{Single vortex line}
\label{sec:1v}

In this Section we analyze the properties of a single vortex line oriented along the $z$ direction 
and located at the center of the trap. We present the results for its core size, its energy and its 
angular momentum across the superfluid-to-supersolid transition. Furthermore  we show that, similarly 
to the case of usual superfluids \cite{recati2001,dalibard2001}, the nucleation of vortices in a 
rotating trap, is dictated by the quadrupole deformation of the superfluid component. 

\subsection{Vortex core structure}

The structure of the vortex core in superfluids is deeply connected to a length called healing length. 
In condensates with only contact interactions, the healing length is computed as the half width at half 
maximum of the wave function. Keeping the same definition also the dipolar gases, seminal papers 
already studied in detail the dependence of the healing length on the scattering length for dipolar 
gases without the LHY correction~\cite{Abad2009,odell2007a}. In these works, it has been shown that 
the healing length of the vortex increases by increasing $\varepsilon_{\rm dd}$ until the gas collapses.

As mentioned above, quantum fluctuations prevent the collapse, and the supersolid phase emerges in the 
trapped geometry at higher values of $\epsilon_{dd}$. In Fig.~\ref{figcore} we report the healing length 
as the system goes from the superfluid to the supersolid regime, by solving in imaginary time the 
Gross-Pitaevskii equation in a rotating frame with angular frequency $\Omega$. We find that the healing 
length keeps increasing till the transition point, after which a non-monotonic and irregular behavior is 
observed. At the transition point, there appears to be a jump. In the supersolid phase, the healing 
length does not show a monotonic behaviour, but it remains roughly constant. As noticed already in 
\cite{roccuzzo2} indeed in the supersolid phase the vortex core size is of the same order of the 
peak density distance, which implies that the vortex core is no longer characterized only by atom-atom 
interactions, but it is deeply modified by the crystal structure. 
The healing length for the non-dipolar case is also shown for comparison.

\begin{figure}[]
\centering
\includegraphics[width=\columnwidth]{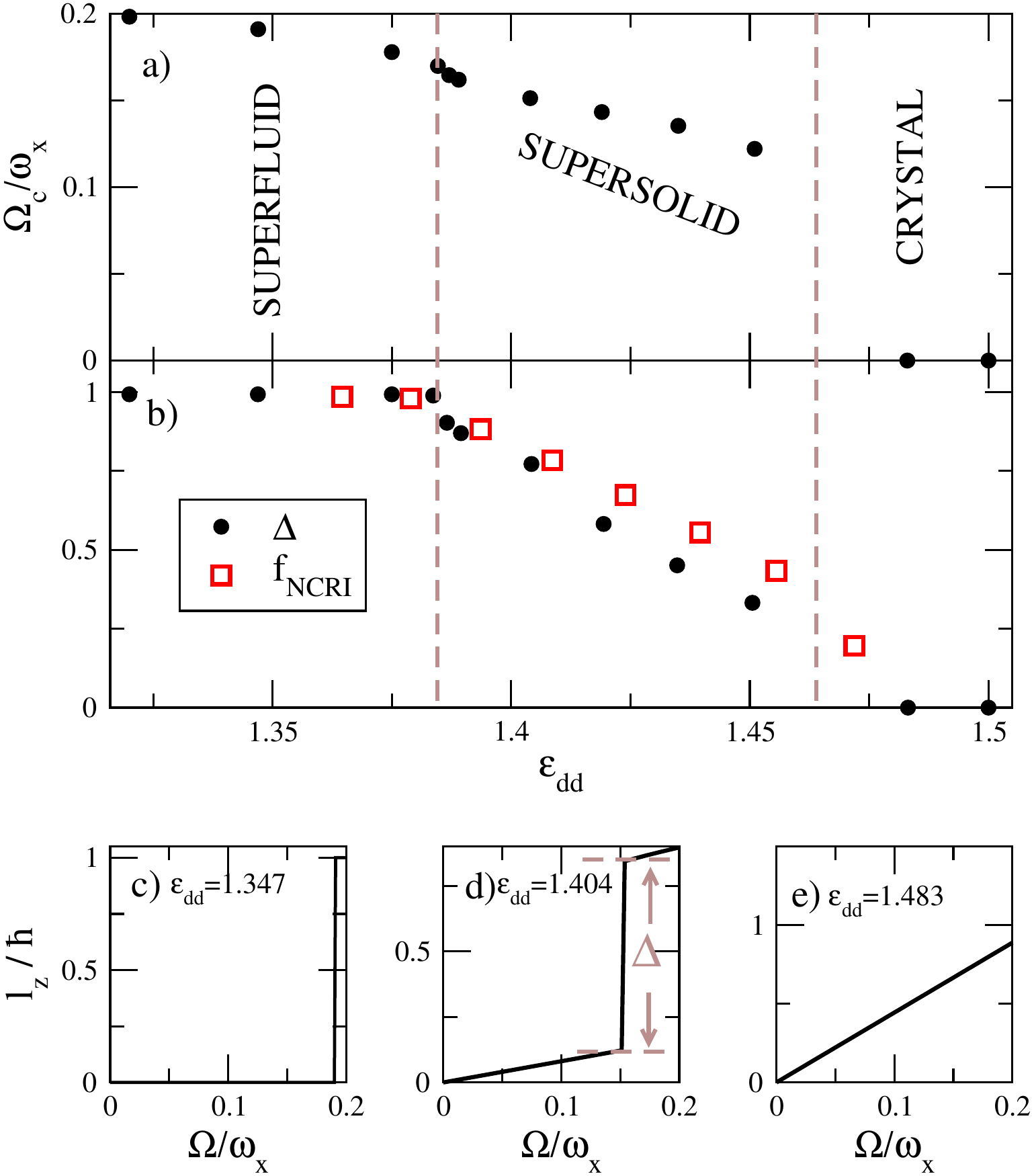}
\caption{a) Critical rotation frequency as a function of $\varepsilon_{\rm dd}$. b) Jump of the 
angular momentum $\Delta$ and nonclassical momentum of inertia fraction $f_{\rm NCRI}$ as a function 
of $\varepsilon_{\rm dd}$. Angular momentum as a function of the rotation frequency $\Omega$ for c) 
$\varepsilon_{\rm dd}=1.347$, d) $\varepsilon_{\rm dd}=1.404$ and  e) $\varepsilon_{\rm dd}=1.483$, 
respectively in the superfluid, supersolid and crystal regimes. The other parameters are the same of Fig.~\ref{figcore}, i.e., $N=40000$ 
atoms and trapping frequencies$\omega_{x,y,z}= 2\pi\times (60,60,120)$ Hz. }
\label{crit_rotation}
\end{figure}

\subsection{Critical rotation frequency}

The single vortex line is energetically stable only above a certain angular frequency 
$\Omega_c$, which makes the energy in the rotating frame of the system with the vortex lower than the 
energy without the vortex. In Fig.~\ref{crit_rotation}(a) we show the numerical value of the critical rotation frequency for a stable 
vortex line in the trap center as a function of $\varepsilon_{dd}$ across the whole phase diagram from the 
superfluid to the supersolid and to the droplet crystal regime. There are already several works 
\cite{Abad2009,odell2007a,Cai2018} accounting for the dependence of the critical rotation frequency as a 
function of $\varepsilon_{dd}$ in dipolar condensates without including quantum fluctuations. In that 
case, $\Omega_c$ increases with $\varepsilon_{dd}$, reaching a maximum for $\varepsilon_{dd}=1$, and 
decreasing for larger values until the collapse is achieved. Thanks to the inclusion of the beyond 
mean-field term in the Gross-Pitaevskii equation one can go beyond the mean-field collapse, eventually 
entering the supersolid phase. We find that after the maximum is reached, the critical frequency keeps 
decreasing, showing a rather small jump at the transition to the supersolid regime, and continues 
decreasing until the crystal phase is reached.

\subsection{Angular momentum carried by a vortex}

In a fully superfluid system, the angular momentum per particle carried by a vortex line is $\hbar$. 
In a partially superfluid system, this value should instead become smaller. The angular momentum 
carried by the vortex corresponds to the jump $\Delta \hbar$ in the angular momentum per particle at 
$\Omega=\Omega_c$ by increasing the angular velocity from below to above the critical value. We have 
determined such a jump across the whole zero-temperature phase diagram of the trapped dipolar gas 
and the value of $\Delta$ is reported in Fig.~\ref{crit_rotation}(b). As intuitively expected, we 
find that $\Delta=1$ in the superfluid phase, it decreases monotonically in the supersolid phase 
and eventually becomes zero in the  droplet crystal phase. For completeness, the angular momentum 
per particle as a function of $\Omega$ is reported in Figs.~\ref{crit_rotation}(c), (d) and (e), 
corresponding to the superfluid, supersolid and droplet crystal regime, respectively.

It is interesting to analyze the connection between $\Delta$ and the superfluid fraction of the system, 
through the jump in the moment of inertia. Such a jump is only due to the presence of a superfluid part 
in a supersolid, and therefore $\Delta$ itself is a natural quantity to evaluate the global superfluidity 
of the system. Another very relevant quantity to characterise the superfluidity of finite systems 
is the nonclassical rotation of inertia fraction that we studied in detail for the dipolar supersolid gas 
in Ref.~\cite{roccuzzo2}. The nonclassical rotation of inertia fraction is given by  
\begin{equation}
f_{\rm NCRI}=1-\Theta/\Theta_{\rm rig}\,,
\end{equation}
where $\Theta$ is the moment of inertia of the system and $\Theta_{\rm rig}$ is its rigid body value. 
As pointed out by Leggett \cite{Leggett1970} in cylindrical annulus, $f_{\rm NCRI}$ coincides with the 
superfluid fraction $f_s$. In Fig. 1(b) we compare $f_{\rm NCRI}$ and $\Delta$. In the superfluid phase, 
they are both equal to $1$, i.e., the whole system is superfluid. In the supersolid region they start 
deviating from each other with $f_{\rm NCRI}>\Delta$. In the droplet crystal phase $f_{\rm NCRI}$ remains 
finite, while $\Delta=0$. The reason is due to the fact that each density peak (droplet) in the crystal 
phase is superfluid by itself, increasing therefore the value of the nonclassical rotation of inertia. 
On the other hand $\Delta$ only accounts for the superfluid component participating to the rotation due to 
the presence of a vortex phase in the order parameter.

\subsection{Quadrupole Instability and Vortex Nucleation}

\begin{figure}[]
\centering
\includegraphics[width=\columnwidth]{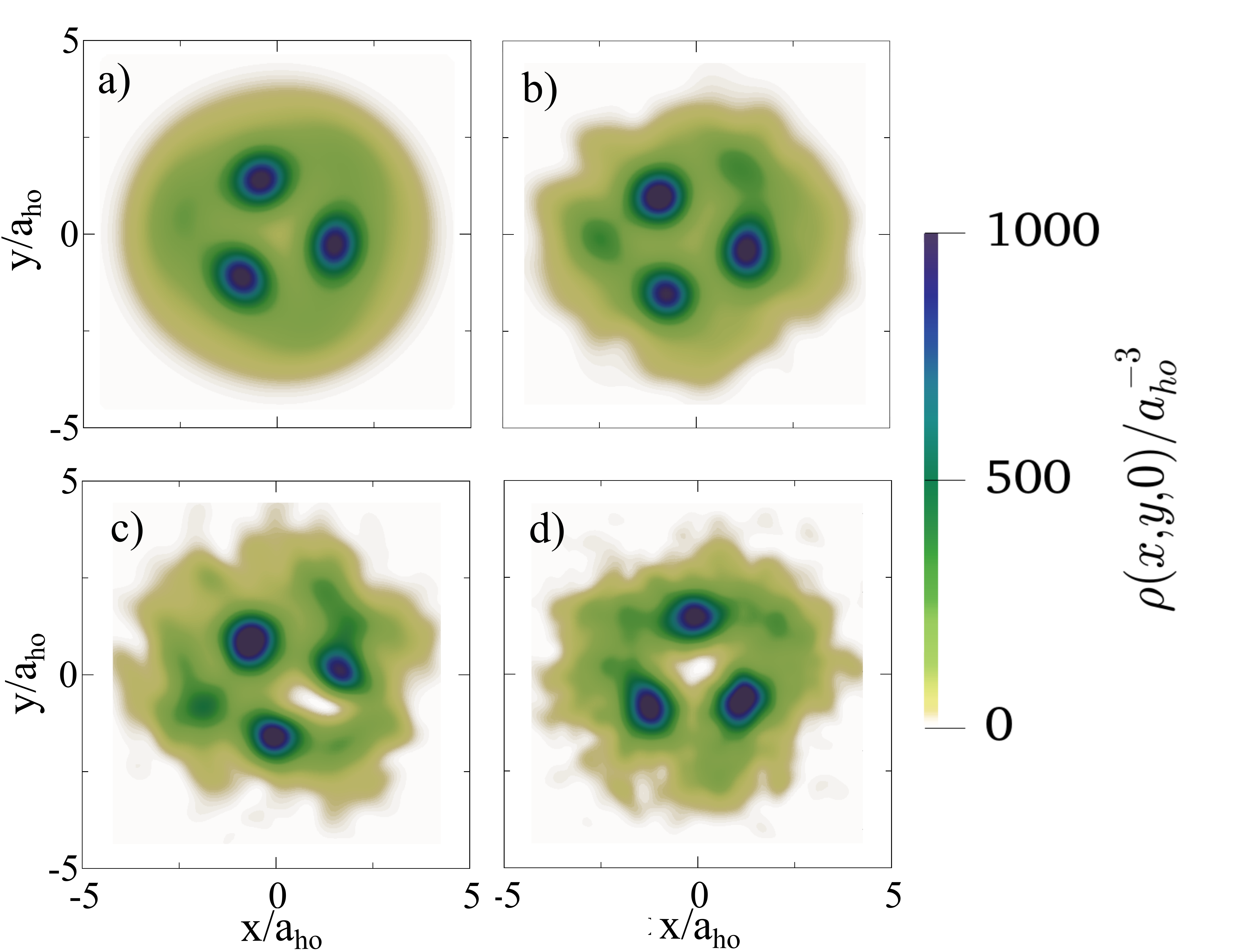}
\caption{In-situ density profiles along the $z=0$ plane showing the nucleation of a 
vortex in a gas of $N=40000$ $^{164}$~Dy atoms in a slightly deformed trap, of frequencies 
$\omega_{x,y,z}=2\pi\times(59.9,60.1,120)$ Hz and $\epsilon_{dd}=1.394$. a) Initial 
preparation of the gas in the supersolid ground state. b) The system is put in rotation 
by the adiabatic introduction of an angular momentum constraint, until the angular 
velocity of $2\pi\times 20$ Hz is reached. The system shows a slight quadrupolar 
deformation in the $z=0$ plane. Several vortices forms at the surface of the system. 
c) The vortices try to penetrate the lattice through the interstitial region between 
the droplets, in order to lower the energy. d) A single vortex finally settles in the
middle of the trap.}
\label{fig1_santo}
\end{figure}

\begin{figure}[t!]
\centering
\includegraphics[width=\linewidth]{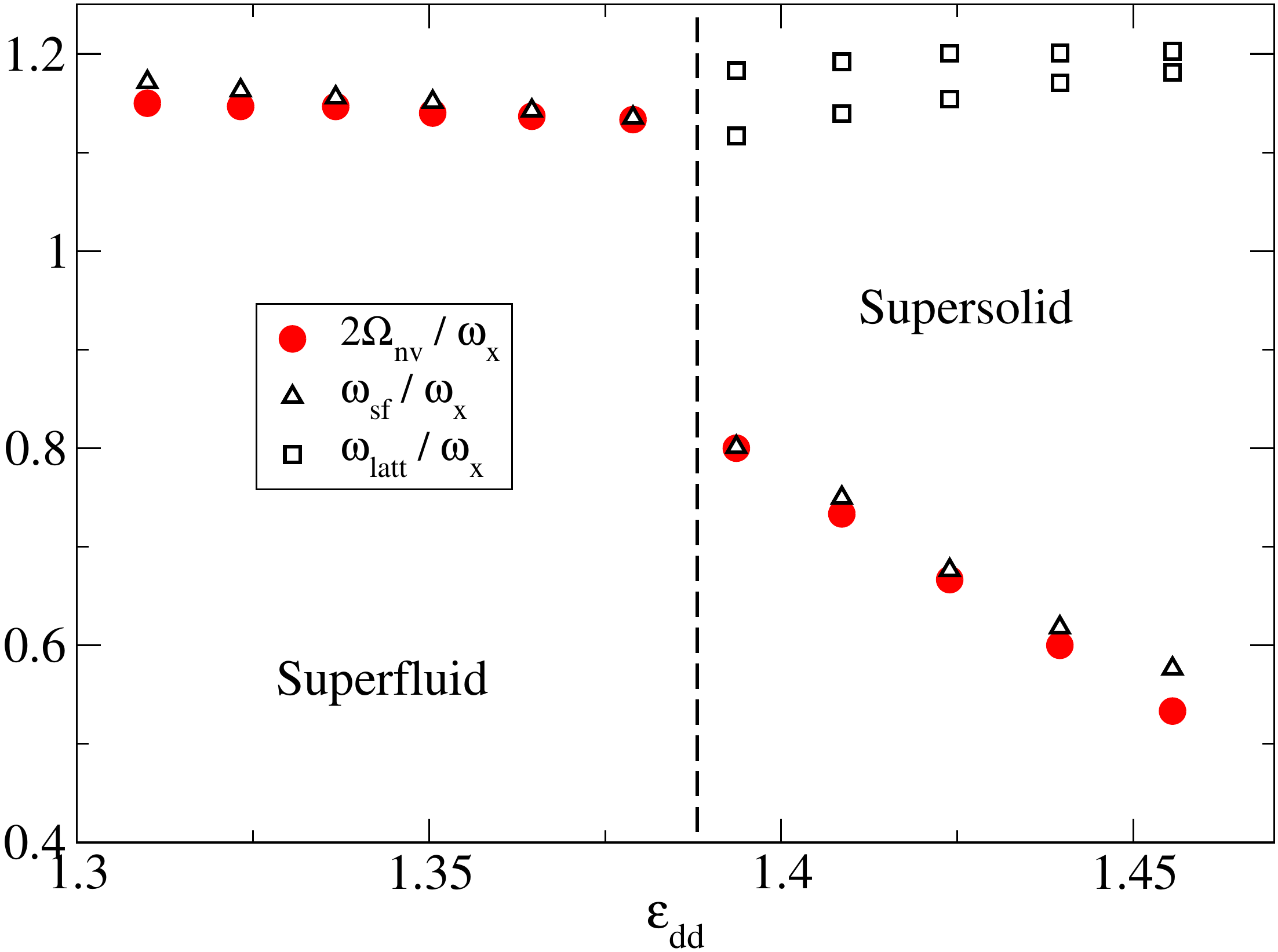}
\caption{Twice the critical frequency for vortex nucleation via the introduction of a 
rotating quadrupolar deformation (red circles), and frequencies of quadrupolar compressional 
modes (empty symbols), as a function of $\epsilon_{dd}$. While in the superfluid phase one 
finds a single mode excited by a sudden quadrupolar deformation (triangles), in the supersolid 
phase one finds three modes, two of which are associated with lattice excitations (empty squares) 
and one with superfluid oscillations (triangles), reflecting the presence of three Goldstone modes 
in an infinite system. The relation $\Omega_c=\omega_{q}/2$, expected for a superfluid (see text), remains 
valid also in the supersolid.}
\label{fig2_santo}
\end{figure}

In this Section we address the problem of vortex nucleation by rotating the harmonic trap, as 
implemented for the first time almost 20 years ago for a standard $^{87}$~Rb BEC~\cite{dalibard2001}. 
The dynamics of vortex nucleation in ordinary (non-dipolar) condensates in a rotating trap has been 
extensively studied. Vortex nucleation is induced by the introduction of a suitable rotating 
deformation of the trap, characterized by a rotation frequency $\Omega$ and deformation parameter 
$\epsilon=(\omega_x^2-\omega_y^2)/(\omega_x^2+\omega_y^2)$.

It turns out that there exists indeed a critical frequency for vortex nucleation $\Omega_{\rm vn}$ 
\cite{dalibard2000,dalibard2001}, which is significantly higher than the one at which a vortex becomes 
energetically favourable. The reason is due to the presence of an energetic barrier~\cite{BecBook2016} 
for the vortex to enter due to the need of creating a density depletion at the vortex position. 
In Refs. \cite{recati2001, sinha2001} it was shown that for rotating harmonic traps, the mechanism of 
vortex nucleation is triggered by the dynamic instability of the quadrupole mode, according to the 
resonance condition
\begin{equation}
\Omega_{\rm vn}=\omega_{q}/2\,,
\label{resfreq}
\end{equation}
with $\omega_{q}=\sqrt{2}\,\omega_{\perp}$ the frequency of the quadrupole mode in the absence of 
rotation \cite{stringari96}. The dynamical instability leads to the spontaneous breaking of the 
cylindrical symmetry of the cloud creating the condition for vortices to be nucleated 
\cite{recati2001, sinha2001}.

Not surprisingly, considering the Gross-Pitaevskii equation without the LHY term, dipolar superfluids 
show the same kind of quadrupole instability \cite{odell2007b,Abad2009}, which could therefore drive 
vortex nucleation, when the resonance condition is satisfied. Notice however that for a dipolar gas, 
the quadrupole frequency $\omega_q$ is not simply given by $\sqrt{2}\omega_\perp$, but it depends on 
the interaction strength and the trapping parameters of the dipolar gas (see, e.g. \cite{odell2007b}). 
Quite remarkably, by direct numerical simulations, we have shown that the critical frequency for the 
vortex nucleation is still given by the resonant condition Eq.(\ref{resfreq}) also when the LHY 
correction is included and even when the system is in the supersolid phase.

We consider $N=40000$ atoms confined in a harmonic trap with frequencies 
$\omega_{x,y,z}=2\pi\times(60,60,120)$ Hz, whose ground state configuration is 
obtained by propagation in imaginary time of Eq.~(\ref{gpequation}). The quadrupole 
mode frequency of the system is obtained by evolving the system in real time under 
the action of a small, sudden quadrupolar deformation of the trap. We find that a 
sudden quadrupolar perturbation in the supersolid phase excites three modes that 
can be associated with the three Goldstone modes expected for an infinite, quasi-2D 
supersolid~\cite{AndreevLifshitz,Macri_SS}: one Goldstone mode associated with the 
spontaneous breaking of the U(1) symmetry responsible for superfluidity and two associated 
with the spontaneous breaking of translational invariance along two directions. The results 
for the quadrupole frequencies are reported in Fig.~\ref{fig2_santo}. They extend to the 
2D case our previous findings for the axial breathing mode of an elongated system~\cite{F2}. 
As expected we find that the lower mode decreases as $\epsilon_{dd}$ is increased, being 
dominated by the (global) superfluidity of the system, which disappears approaching the 
droplet crystal phase. The other two frequencies, dominated by the motion of the crystal peaks, 
increase until saturation in analogy to what we found in~\cite{F2}.

The nucleation of the vortex is studied instead by evolving Eq.~(\ref{gpequation}) in 
real time starting from the ground state by adding the $-\Omega L_z$ term to the 
Hamiltonian~\footnote{Aside from the time scale on which 
the dynamics occurs, the final results and our conclusions do not change if instead of a 
step function we use, e.g., a linear ramp for $\Omega(t)\propto t$}. A very small trap 
deformation ($\epsilon=3.33\times10^{-3}$) to trigger the instability is also added. We 
observe that for any value of $\varepsilon_{dd}$ there exists a critical angular frequency 
$\Omega_{\rm vn}$, such that for $\Omega\ge\Omega_{\rm vn}$ strong cloud deformations occur 
followed by the nucleation of a vortex. Typical snapshots during the time evolution are reported  in Fig.~\ref{fig1_santo}~\footnote{The full movie 
of the vortex nucleation is available as Supplementary Material}, where lengths are in units of the geometric mean of the 
three harmonic oscillator lengths $a_{\rm ho}=\sqrt{\hbar/m}\times(\omega_x\omega_y\omega_z)^{-1/6}$. 

In Fig.~\ref{fig2_santo} 
we report the calculated values of $\Omega_{\rm vn}$ multiplied by the factor $2$ in order 
to make the comparison with the lowest quadrupole frequency more direct and explicit. The 
resonance condition Eq.~\ref{resfreq} considering the (superfluid) lower frequency mode is 
met.

A comment on the time scale for vortex nucleation is due here. Our simulation predicts rather 
long times (of the order of $1$ second) for the vortex nucleation. However, in real experiments, 
noise and thermal effects are expected to trigger the instability on a much faster time scale. 
Larger trap deformations also help in speeding up the nucleation process.

\section{Vortex lattices in dipolar supersolids}
\label{sec:lattice}
\begin{figure*}[t!]
\centering
\includegraphics[width=\linewidth]{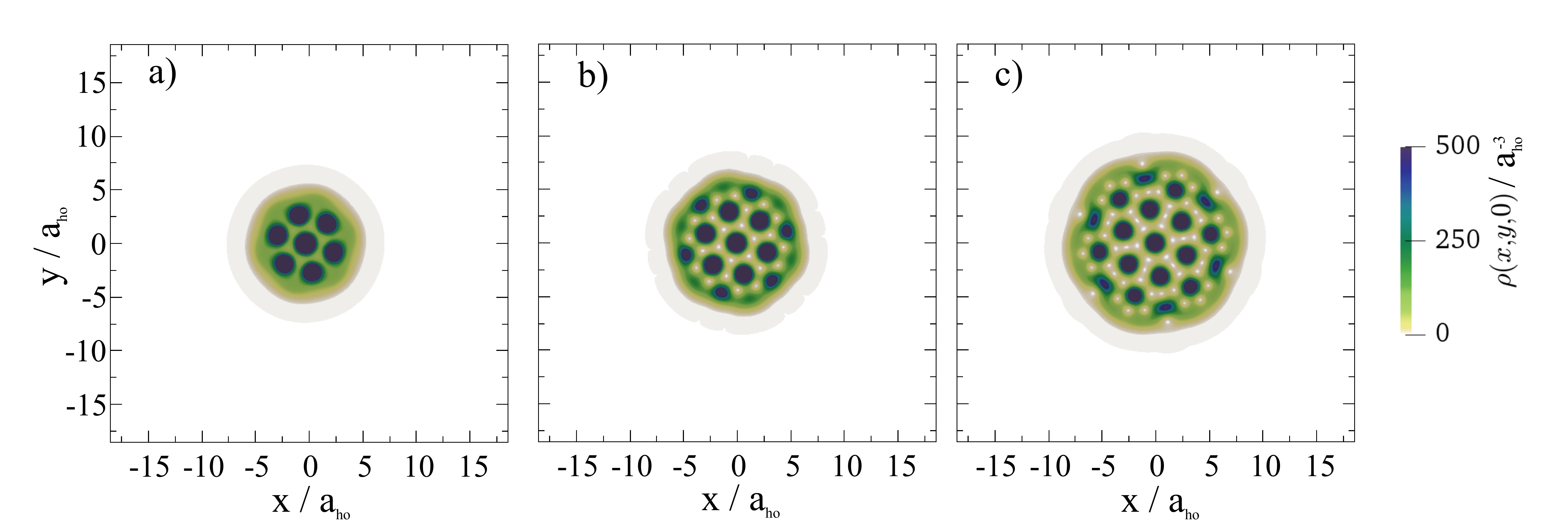}
\caption{Density plots of $N=110000$ $^{164}$~Dy atoms in the supersolid 
phase for $\epsilon_{dd}=1.409$, confined in an harmonic trap with frequencies 
$\omega_z=2\omega_x=2\pi\times 120$~Hz. Panel a): non-rotating gas $\Omega=0$, 
Panel b) $\Omega=2/5\omega_x$ and Panel c) $\Omega=5/6\omega_x$.}
\label{fig_lattice}
\end{figure*}

A superfluid can host more than a single vortex line, if the rotational frequency is large 
enough. Many vortices form, in the $z=0$ plane, a 2D triangular lattice called {\it{Abrikosov 
lattice}}~\cite{Abo-Shaeer476,Cornell2001}. The aim of this Section is to address the question 
of whether and how a dipolar supersolid can host many vortices. The presence of the regular 
(triangular lattice) density modulation can indeed interfere with the formation of the Abrikosov 
lattice, when the intervortex distance (which scales as $\Omega^{-1/2}$) becomes of the same 
order of the distance between the density peaks (the latter being fixed by the roton wave 
vector~\cite{SantosRoton}, which in supersolids is of the order of a few units of the axial 
oscillator length $\sqrt{\hbar/m\omega_z}$).

We have first checked that the triangular Abrikosov lattice persists in the whole superfluid regime, 
also in the presence of the dipolar interaction and LHY term, till the transition to the supersolid 
phase~\footnote{It is important to notice that in spite of the fact that the vortex core can be much 
larger in the dipolar case than in the more standard Bose condensates with only contact interaction 
(see Fig.~\ref{figcore}), the number of vortices is independent of their size, being fixed by the 
value of the rotation frequency. }.

In the supersolid phase, it is energetically favourable to accommodate the vortices in the low density 
regions. Therefore when the vortex distance is of the same order of the distance between the density 
peaks, there is a competition between the natural tendency of vortices forming the triangular Abrikosov 
lattice and the vortices occupying the valleys of the supersolid density. In the panel (b) of 
Fig.~\ref{fig_lattice}, we show the numerical result of the imaginary-time evolution of the extended 
Gross-Pitaevskii equation in a rotating frame for large enough angular velocities. We obtain that the 
vortices are pinned by the minima of the supersolid density modulations forming -- for the choosen 
value $\Omega=2/3~\omega_x$ -- a honeycomb lattice. It is instructive to compare it with the system 
for $\Omega=0$, reported in panel (a). The increase of the cloud's radius (and the density reduction) 
due to the centrifugal potential allows a larger number of peaks to be hosted in the system.

It is important to notice that in the literature the pinning of Abrikosov vortex lattices in Bose gases 
has been addressed by a number of authors considering an underlying (square or triangular) rotating 
optical lattice \cite{BigelowPinning,Reijnders2005,NikuniPinning,MuellerPinning}, with a nice early 
experimental demonstration of the transition from the ``natural" Abrikosov lattice to the pinned vortex 
lattice by Eric Cornell and coworkers \cite{EricPinning}. We remind, however, that in our case, the 
structure of the density modulation is not imposed by an external potential, but is due to the 
spontaneous breaking of translational symmetry, yielding the supersolid phase.

The pinned honeycomb lattice persists as long as $\Omega$ is not too large for the intervortex 
distance to become smaller than the period of the supersolid lattice. By further increasing $\Omega$ 
we find that the vortices are hosted in the low density regions surrounding the droplets as shown 
in the panel (c) of Fig.~\ref{fig_lattice} for $\Omega=5/6~\omega_x$.

\section{Expanding a Supersolid with vortex lines}

\begin{figure}[]
\includegraphics[width=\columnwidth]{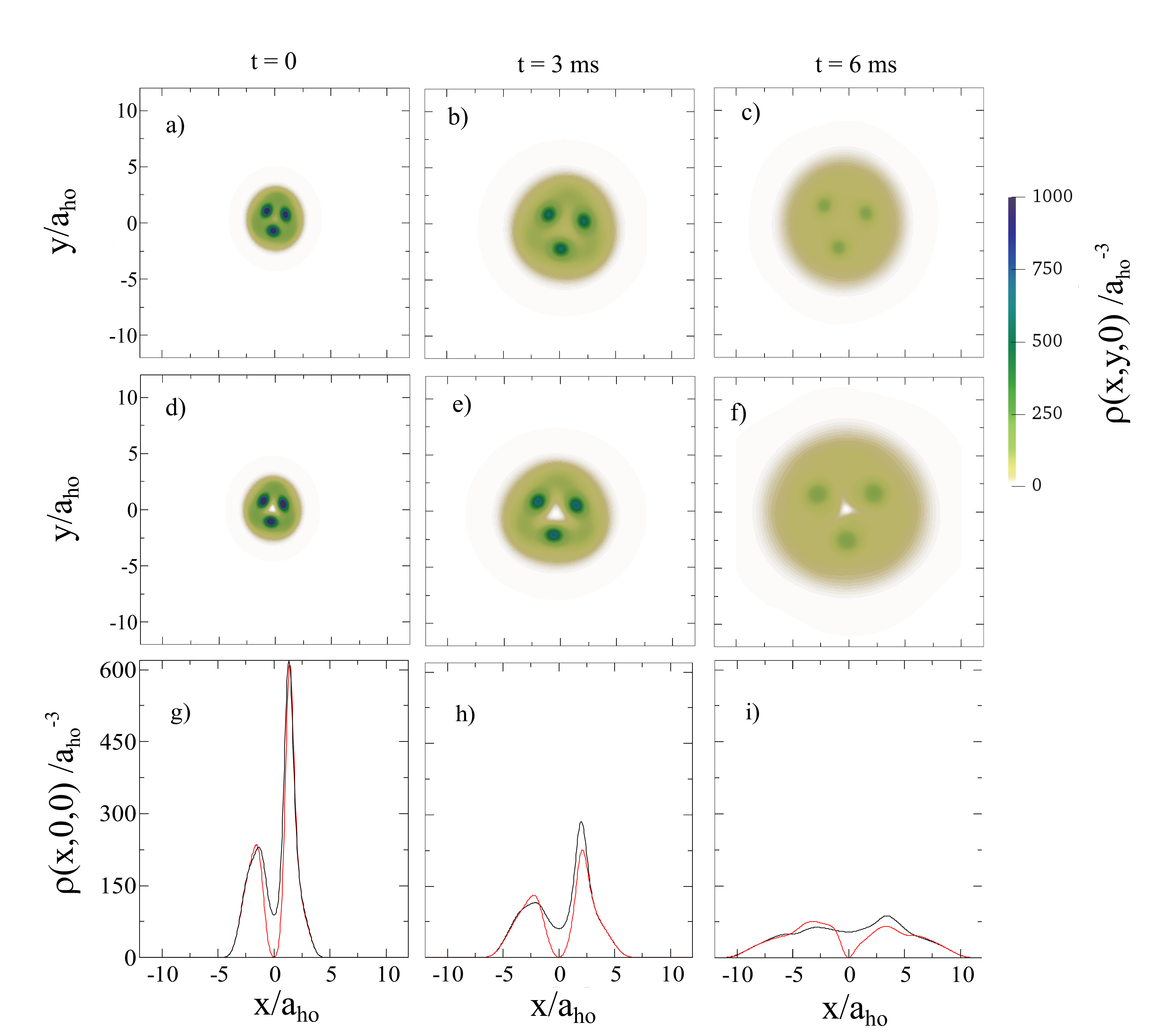}
\caption{In-situ density profiles along the $z=0$ plane (panels (a) to (f)), and cuts along 
the $x$-axis (panels (g-i)) of an expanding dipolar supersolid in the absence (a-c) and in 
presence (d-f) of a vortex. The initially prepared ground state in the presence (d) of a 
vortex has been obtained for $\Omega=2\pi\times20$ Hz. Panel (g) shows the corresponding 
density cut along the $x$-axis. The red (black) line corresponds to the case with (without) 
vortex. Panels (h) and (i) show corresponding cuts along the $x$-axis. The other parameters 
are the same as in Fig. \ref{fig1_santo}. A full movie of the expansion is available as 
Supplementary Material.}
\label{fig3_santo}
\end{figure}

\begin{figure}[]
\includegraphics[width=\columnwidth]{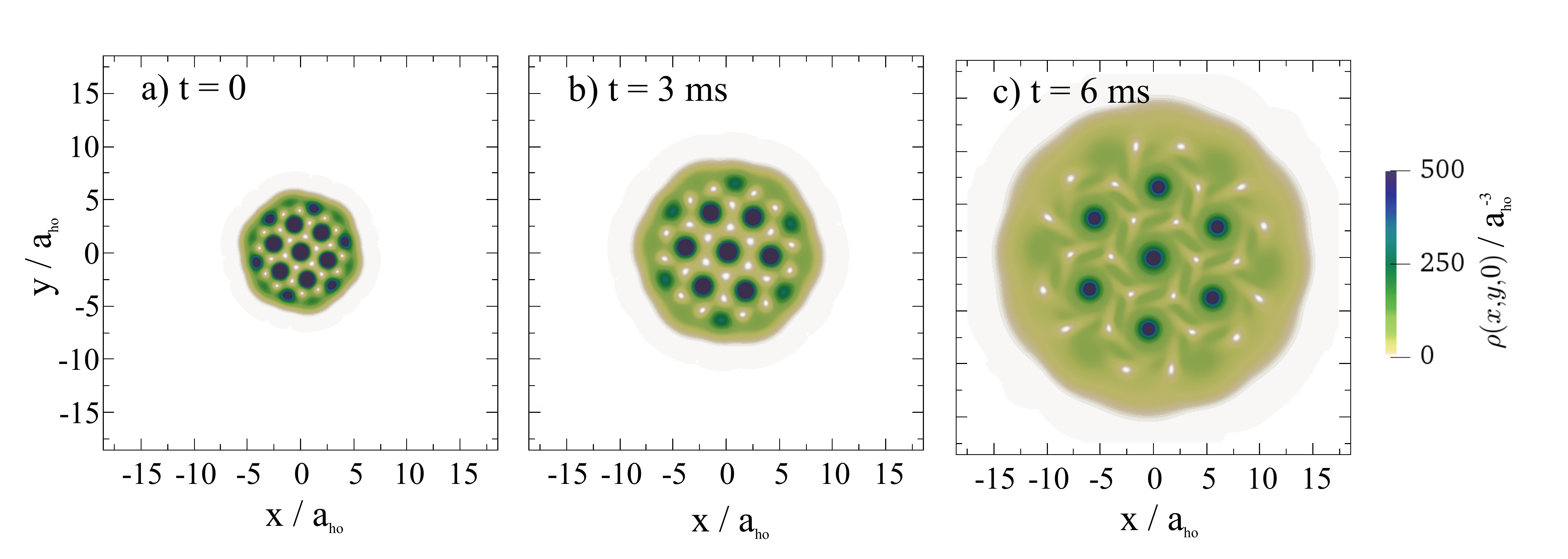}
\caption{Expansion of a dipolar supersolid in the presence of the vortex lattice reported in 
Fig.~\ref{fig_lattice} b). Panel (a) the initially prepared ground state. This state is evolved 
in real time, after switching off the radial confinement. Panel (b) and (c) show the $z=0$ density 
profile at two later times.
A full movie of the expansion is available as Supplementary Material.}
\label{fig4_santo}
\end{figure}

The previously reported results considered the possibility of addressing the system in-situ. Here we 
briefly discuss the effect of letting the cloud to expand, i.e., after switching off the trap in the 
transverse ($z=0$) plane, in order to image the system with a better space resolution. We consider both 
the single and the many-vortex case.

Figure \ref{fig3_santo} shows the density profiles of a dipolar supersolid with and without a vortex 
line at two different times after the removal of the trap~\footnote{The full dynamics of the vortex 
expansion is made available in the Supplementary Material}. The ratio between the peak density and 
the central density, in the absence of the vortex, is less than $10$ and it further decreases during 
the expansion, the minimum of the  density remaining of the same order as that of an ordinary 
superfluid. Thus, with our choice of parameters, a good imaging system could easily identify the 
presence of the vortex in the center of the trap.

For the high angular frequency case, when many vortices appear, we consider the most interesting case, 
when the vortices form a honeycomb lattice, as in Figure~\ref{fig_lattice} b). The expansion at two 
different times after switching off the transverse confinement is reported in 
Fig.~\ref{fig4_santo}~\footnote{The full dynamics of the vortex lattice expansion is made available 
in the Supplementary Material}. In particular, we notice that the geometry of the two lattices remains 
unchanged during the expansion, paving the way for the possible direct observation of the frustration 
of the vortex lattice.

Concerning the latter case, while finishing the present work, we became aware of a very recent 
work~\cite{ancilotto2020vortices} discussing a novel protocol to produce a vortex lattice in a 
large supersolid cloud and exploring the following dynamics of the expansion.

\section{Conclusions}
\label{sec:conc}

In this work we have provided a first comprehensive study of quantized vortex lines in the supersolid 
phase of an ultracold dipolar gas rotating in the plane orthogonal to the polarization direction of 
the dipole moment of the atoms. The analysis has been carried out by means of the generalised 
Gross-Pitaevskii equation (Eqs.~\ref{gpequation}-\ref{gpequation2}), including LHY corrections for 
the stabilisation of the supersolid phase. 

We have studied in detail the stationary properties and the nucleation dynamics of a single vortex 
line hosted in the center of the cloud. 
We have found  that the width of the vortex lines, close and in the supersolid phase, is significantly larger than in usual condensates interacting with contact forces. The width is large enough to be  likely  directly imaged with available in-situ techniques. The critical rotational frequency for the energetic stability of a vortex has been found to decrease by increasing the dipolar interaction strength close 
to and into the supersolid phase.
 The angular momentum carried by a vortex is smaller than $\hbar$ in 
the supersolid phase and approaches zero by approaching the crystal droplet phase, due to the reduction 
of the superfluid density of the cloud. Remarkably we could show that the nucleation of a vortex is 
always triggered by the softening of the lowest quadrupole mode frequency (see Fig.~\ref{fig2_santo}), 
as in standard Bose-Einstein condensates with contact interaction.   

At large enough angular velocity we have addressed the problem of the spatial arrangement of many vortices.
The supersolid phase forces the vortices to be pinned in the density valleys. In particular we have shown 
that vortices arrange in a regular honeycomb lattice when the intervortex distance is of the same order 
of the solid periodicity (Fig.~\ref{fig_lattice} (b)), the novel supersolid vortex structure being preserved during the free expansion, following the release of the trap.

\section*{Acknowledgement}
Useful discussions with
Franco Dalfovo, Giacomo Lamporesi and with the members of the Ferlaino's Dipolar Quantum Gas group are acknowledged. This project has received funding from the
European Union's Horizon 2020 research and innovation
programme under grant agreement No. 641122 ``QUIC",
from Provincia Autonoma di Trento, the Q@TN initiative and the FIS$\hbar$ project of the Istituto Nazionale
di Fisica Nucleare.


\bibliography{biblio-inertia.bib}

\end{document}